\documentclass[useAMS,usenatbib]{mn2e}
\usepackage{graphicx}
\usepackage{natbib}

\title[Cusp regrowth in dwarf galaxies]{Under the sword of Damocles: plausible regeneration of dark matter cusps at the smallest galactic scales}
\author[Laporte  \& Pe\~narrubia]{ \parbox[t]{\textwidth}{
Chervin F. P. Laporte$^{1,2}$ \& Jorge Pe\~narrubia$^{3}$}
\\
$^{1}$ Max Planck Institute for Astrophysics, Karl-Schwarzschild-Strasse 1, 85740 Garching, Germany\\
$^{2}$ Department of Astronomy, Columbia University, 550 West 120th Street, New York, NY, 10027, U.S.A\\
$^{3}$ Institute for Astronomy, The University of Edinburgh, Royal Observatory, Blackford Hill, Edinburgh EH9 3HJ, UK \\
}
\begin{document}
\date{}
\pagerange{\pageref{firstpage}--\pageref{lastpage}} \pubyear{2011}
\maketitle
\label{firstpage}
\begin{abstract}

We run controlled N-body experiments to study the evolution of the dark matter (DM) halo profiles of dwarf galaxies driven by the accretion of DM substructures. Our initial conditions assume that supernova feedback erases the primordial DM cusps of haloes with masses $10^{9}-10^{10} \rm{M_{\odot}}$ at $z=0$. The orbits and masses of the infalling substructures are borrowed from the {\it Aquarius} simulations. Our experiments show that a fraction of haloes that undergo 1:3 down to 1:30 mergers are susceptible to reform a DM cusp by $z\approx 0$. Cusp regrowth is driven the accretion of DM substructures that are dense enough to reach the central regions of the main halo before being tidally disrupted. The infall of substructures on the mean of the reported mass-concentration relation and a mass ratio above 1:6 systematically leads to cusp regrowth. Substructures with 1:6 to 1:8, and 1:8 to 1:30 only reform DM cusps if their densities are one and two-sigma above the mean, respectively. The merging timescales of these dense, low-mass substructures is relatively long $(5-11 \rm{Gyrs})$, which may pose a timescale problem for the longevity of DM cores in dwarf galaxies. These results suggest that a certain level of scatter in the mass profiles of galactic haloes acted-on by feedback is to be expected given the stochastic mass accretion histories of low-mass haloes and the diverse star formation histories observed in the Local Group dwarves.
\end{abstract}
\begin{keywords} galaxies: formation - galaxies: evolution - galaxies: dwarf
\end{keywords}

\section{Introduction}
In the standard $\Lambda$CDM cosmological paradigm, dark matter (DM) constitutes 25 percent of today's total energy density. It is thought to be made up by collisionless particle(s) which at very early times are distributed in a cold configuration (i.e. with zero velocity dispersion). This state is unstable and leads to gravitational clustering, which is typically studied with the aid of cosmological N-body simulations \cite{Efsthathiou1985}. Over the last thirty years, cosmological N-body simulations have enabled us to predict the large-scale structures \citep{Davis1985}, the structure of DM haloes and their associated substructure content from the scales of galaxy clusters to that of the Milky Way - MW - \citep{Dubinski1991, Navarro1996c, Moore1999}.  One key prediction from the CDM paradigm is the existence of completely dark substructures lingering in all galactic haloes, also in those of dwarf galaxies. High-resolution simulations of MW-like DM haloes can resolve structures with mass ratios above 1:100 within dwarf galaxies \citep{Springel2008}, although the internal dynamics of these objects cannot be fully resolved (e.g. their density profiles do not reach a convergent result). A large body of observational data suggest that dwarf galaxies (that is, low-surface brightness and dwarf spheroidals) inhabit DM haloes with cored density profiles \citep{Oh2011, Walker2011, Amorisco2012}. Although this continues to be a lively debated topic (e.g. \cite{Strigari2014}) the so called core-cusp problem currently motivates alternative DM models \citep{Spergel2000, Bode2001, Vogelsberger2012, Rocha2013, Vogelsberger2014, Lovell2014}. Within the CDM paradigm, baryonic feedback \citep{Navarro1996, Governato2010, Zolotov2012, Dicintio2014, Madau2014} has been proposed as a plausible solution to the inference of DM cores on the smallest galactic scales. Strong fluctuations in the gravitational potential driven by supernova explosions may flatten the inner density profile \citep{Pontzen2012}. However, the invoked efficiencies (or coupling) between the kinetic energy liberated during SNe explosions and the surrounding gas are highly uncertain \citep{Garrison-Kimmel2013}. Simple analytical arguments as well as hydrodynamical simulations show that this mechanism is specially efficient at high redshifts\citep{Penarrubia2012, Amorisco2014, Davis2014}.

 One problem remains, however: N-body simulations of structure formation predict the presence of a myriad of DM substructures surrounding dwarf galaxies, and some of them may get close enough to the core to regrow a cusp. \cite{Dekel2003} presented the first experiments of cusp regrowth and showed that for typical CDM orbits 1:10 mergers may regrow a DM cusp within a Hubble time. However, the set-up of these simulations was fairly generic. To date, not attempt to test these results on the scales of dwarf galaxies has been made. Here we use the merger trees of some of the massive satellites formed in the Aquarius runs ($10^{9}-10^{10} \rm{M_{\odot}}$ at $z=0$) in order to identify potential merger candidates and extract their orbits and mass structure at their time of infall. These systems are evolved at an enhanced resolution in order to assess whether cusp-regrowth is plausible mechanism on the scales of dwarf galaxies. To this end we assume that baryonic feedback flattens the primordial DM profile of these systems at high-redshifts. Current cosmological hydrodynamical simulations do not reach the required resolution to follow this process in detail, as the radius of convergence of low-mass haloes around MW-like galaxies is $\sim 1 ~\rm{kpc}$ (as reported in \cite{Zolotov2012}), which is larger than the typical scale radii of DM substructures (see below). This letter is organised as follows. Section 2 discusses our numerical set-up and implementation. We present our results in section 3 and discuss them and conclude in section 4.

\section{Numerical methods and simulations}
We use merger trees from six of the most massive satellites in a MW-like DM halo from the {\it Aquarius} simulations suite, Aq-A-2 which has been run with much more accurate time slicing than any other runs; the details of the simulations can be found in \citep{Springel2008}. The mass and softening resolutions used were $m_{p}= 1.4 \times10^{4} \,\rm{M_{\odot}}$ and $\epsilon= 66 \, \rm{pc}$.
\begin{figure}
\vspace*{0cm}\includegraphics[width=0.45\textwidth,trim=0mm 0mm 0mm 0mm,clip]{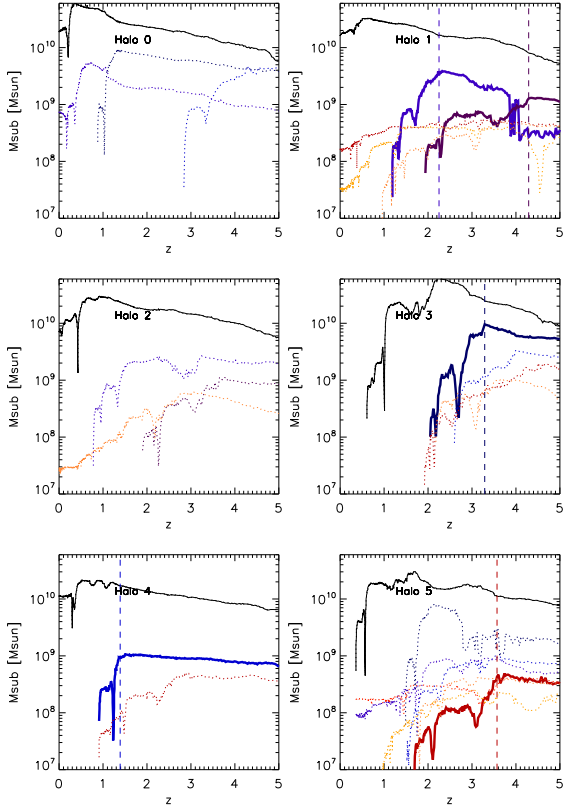}
\caption{Mass evolution of the 6 Aquarius host dwarf DM haloes and their most massive satellites reaching a distance of less than 10 kpc during their orbital history. The black lines in each panel denote the mass evolution of the host dwarf halo. The thick coloured lines represent the subhalos which have been selected for re-simulations to study cusp regrowth. These include a variety of 1:3, 1:6 down to 1:30 mergers. The dashed vertical lines represent the infall redshifts for which we also define the apocentres for the host-subhalo merger re-simulations.}
\end{figure}
For the idealised runs our haloes are modelled as \cite{Dehnen1993} profiles:
\begin{equation}
\rho=\frac{(3-\gamma)M}{4\pi}\frac{a}{r^{\gamma}(r+a)^{4-\gamma}},
\end{equation}
where a is the scale radius, $M$ the total mass and $\gamma$ is a constant in the interval $[0,3)$. The main dwarf hosts are re-modelled with cored
profiles (with $\gamma=0$) while the substructures are re-modelled as \cite{Hernquist1990} profiles (i.e. with $\gamma=1$). The next step is to set a scale radius for the host and its substructures. Given the recorded value of virial mass $M_{200}$ in the simulations, the concentration parameter $c\equiv r_{200}/r_{s} =c(M,z)$ is computed using the empiric relation of \cite{Prada2012} in order to set the scale radius $r_{s}$. We then calculate the radius $r_{-2}$ where the slope of the density profile equals $\gamma(r)=-d\log(\rho)/d\log(r)=2$ for the Dehnen models. This radius corresponds exactly to $a=2\times r_{s}\equiv a_{s}$ for $\gamma=1$, and $a=r_{s}\equiv r_{c}$ for $\gamma=0$. We also need to relate $M_{vir}\equiv M_{200}$ to the mass $M$ of the Dehnen models. For a Hernquist profile
 $M=M_{vir}/f(c)*27/16$, and  $M=M_{vir}/f(c)*4/3$ for $\gamma=0$, where $f(c)=\ln (1 + c) + c/(1 + c)$. In a study on DM profiles formed in cosmological simulations, \cite{Jang2001} show that NFW and Hernquist profiles provide equally good fits to the inner regions of subhalos, although the Dehnen profile is too steep at large radii, which plays a negligible role in our study. N-body models for the dwarf galaxy halo and its substructures are generated through Monte-Carlo sampling the isotropic distribution functions that yield the density profile given in equation (1) using a rejection-acceptance method, as in \citep{Kuijken1994}.

For clarity, we refer to a dwarf galaxy halo as the ``host'' and to the accreted substructures as ``subhalos''. One caveat of idealised simulations is the lack of smooth mass accretion. For every merger event, we generate an N-body realisation of the substructure in dynamical equilibrium with an initial mass $M_{200}$ set at the time of accretion. Subsequently, we place the dark subhalo on an orbit extracted from the merges tree. Note that by keeping the host halo mass fixed our models tend to overestimate the dynamical time scales of the subhalo orbits. Mass ratios $q$,  peri- to apo-centre $r_{a/p}$ ratios (proxies for orbital eccentricities), orbital apocentres $r_{a}$, scale radii of hosts and subhalos $a\equiv r_{c}, a_{s}$ and redshifts of accretion $z_{i}$ are listed in Table 1. Figure 1 shows the mass evolution of the subhalos in our sample, as well as that of their hosts. The subhalos chosen to be simulated with an enhanced resolution are shown by the thick coloured lines. From this Figure it is evident that most accretion events occur well before infall of the host dwarf into their MW-like parent haloes. It appears therefore reasonable to run the dwarf halo models in isolation, that is, omitting the MW's tidal field. We do not simulate the evolution of all substructures appearing in this plot, as many of them have mass-ratios that are too small for dynamical friction to be relevant and do not contribute to the evolution of the host inner density profile. Also, note that the merger of substructures with mass ratios between 1:3 and 1:5 (see \cite{Deason2014} for more quantitative evaluation of this) typically occur at $z \la 5$. If we believe that cores are formed at high redshifts, as suggested by hydrodynamical simulations and analytical arguments \citep{Pontzen2012,Penarrubia2012}, such massive substructures are bound to sink to the centre of the host through dynamical friction and reform a DM cusp. 

To inspect this issue in more detail we run controlled N-body simulations of a number of the merger events selected from Figure~1 and summarised in Table~1. Note that our sample includes subhalos with very similar mass-ratios and/or apo-to-peri ratios (e.g. Halo 2 accretes substrucutures on similar orbits and infall times as those in Halo 1, as shown in Figure 1), so we chose to run only a representative subsample of all the cases shown in Table~1.  A key aspect of our study focuses on the role played by the inner structure of the accreted subhalos. This is done by exploring the scatter in the mass-concentration relation $c(M,z)$ of the DM substructures, whose dispersion is $\sigma_{log_{10}(c)} \sim0.1$ apparently independent of redshift or mass \citep{Neto2007}. As shown below, the mean density of a subhalo determines the survival time-scale and where the stripped material is deposited. To gauge this effect, for every accretion event we adopt concentrations $c =<c>+1\sigma_{c}, 2\sigma_{c}, 3\sigma_{c}$, which cover a fair range of the expected values within CDM paradigm (namely, these models correspond to 30, 13 and 1 percentile of the lognormal distribution of concentration at fixed halo mass). The particle mass used is $m_{p}=10^{4} \rm{M_{\odot}}$ and softening of $\epsilon\sim 50 \rm{pc}$, roughly corresponding to the inter-particle spacing in the realisations considered. All simulations were tested for convergence by running simulations with 10 times a higher mass resolution than quoted (with $m_{p}=10^{3} ~ \rm{M_{\odot}}$ and $\epsilon=30 ~ \rm{pc}$).

\begin{table*}
\centering
\begin{minipage}{160mm}
 \begin{tabular}{@{}llrrrrlrlrlrlrlrlrlrlrlr@{}}
 \hline
Run & $q$ & $M_{v_{h}}$ & $M_{v_{s}}$ & $r_{v_{h}}$ & $r_{c}$ & $a_{s}$ & $M_{h}$ & $M_{s}$ & $\delta_{\rho}$ & $r_{a/p}$ & $r_{a}$ & $z_{i}$ &  $Q_{g}$ & $t_{s}$ \\
&  & $ 10^{10}\rm{M_{\odot}}$ & $ 10^{10}\rm{M_{\odot}}$ & $ kpc $ & kpc & kpc & $10^{10} \rm{M_{\odot}}$ &$10^{10} \rm{M_{\odot}}$ & & kpc & kpc &  Gyrs \\
 \hline
3.4.       & 1:3 & 2.5 &  1.0 & 21 & 5.1 & 6.2 & 3.9 & 1.9 & 0.18 & 14 & 44 & 3 & y & 2.0 \\
1.2.       & 1:6 & 1.7 &  0.3 & 25 & 5.0 & 5.2 & 2.5 & 0.50 & 0.12 &4.0 & 40 & 2  & y & 5.0 \\
1.2.1s  & 1:6 & 1.7 &  0.3 & 25 & 5.0 & 4.0 & 2.5 & 0.42 & 0.22 &4.0 & 40 & 2  & y & 4.6 \\
1.2.2s  & 1:6 & 1.7 &  0.3 & 25 & 5.0 & 3.2 & 2.5 & 0.37 & 0.38 &4.0 & 40 & 2  & y & 4.3 \\
1.2.3s  & 1:6 & 1.7 &  0.3 & 25 & 5.0 & 2.6 & 2.5 & 0.33 & 0.63 &4.0 & 40 & 2  & y* & 4.3-8.0 \\
1.3.       & 1:8 & 0.8 &  0.1 & 12 & 3.0 & 3.2 & 1.3 & 0.26& 0.11&3.0 & 27 & 4.3  & n & - \\
1.3.1s  & 1:8 & 0.8 &  0.1 & 12 & 3.0 & 2.4 & 1.3 & 0.21 & 0.21 &3.0 & 27 & 4.3  & y & 5.0 \\
1.3.2s  & 1:8 & 0.8 &  0.1 & 12 & 3.0 & 2.0 & 1.3 & 0.18 & 0.31 &3.0 & 27 & 4.3  & y* & 4.6-8.0 \\
1.3.3s  & 1:8 & 0.8 &  0.1 & 12 & 3.0 & 1.6 & 1.3 & 0.16 & 0.54 &3.0 & 27 & 4.3  & y* & 4.4-7.0 \\
4.1.       & 1:20 & 1.7 &  0.1 & 31 & 5.0 & 3.2 & 2.0 & 0.14 & 0.18 &18 & 45 & 1.5  & n & - \\
4.1.1s  & 1:20 & 1.7 &  0.1 & 31 & 5.0 & 2.6 & 2.0 & 0.12 & 0.28 &18 & 45 & 1.5  & n& - \\
4.1.2s  & 1:20 & 1.7 &  0.1 & 31 & 5.0 & 2.1 & 2.0 & 0.10 & 0.45 &18 & 45 & 1.5  & y & 7 \\
4.1.3s  & 1:20 & 1.7 &  0.1 & 31 & 5.0 & 1.6 & 2.0 & 0.09 & 0.92 &18 & 45 & 1.5  & y & 6.5 \\
5.5.       & 1:30 & 1.1 &  0.04 & 15 & 3.7 & 2.3 & 1.8 & 0.08 & 0.12 &5  & 22 & 3.6  & n & - \\
5.5.1s  & 1:30 & 1.1 &  0.04 & 15 & 3.7 & 1.8 & 1.8 & 0.07 & 0.22&5  & 22 & 3.6  & n & - \\
5.5.2s  & 1:30 & 1.1 &  0.04 & 15 & 3.7 & 1.4 & 1.8 & 0.06 & 0.41&5  & 22 & 3.6  & y* & 5.0-11.0 \\
5.5.3s  & 1:30 & 1.1 &  0.04 & 15 & 3.7 & 1.2 & 1.8 & 0.05 & 0.54&5  & 22 & 3.6  & y* & 4.0-10.0 \\
\hline
\end{tabular}
\end{minipage}
\caption[]{Simulations' properties: runs (coded as halo.subhalo.$<c>$+$\sigma_{c}$), mass-ratios ($q$), host ($M_{v_{h}}$) and satellites ($M_{v_{s}}$) virial masses, virial radius of host ($r_{v_{h}}$), scale radius of host ($r_{c}$), scale radius of satellite ($a_{s}$), mass of host ($M_{h}$), mass of satellite $M_{s}$ (both chosen to match the corresponding NFW profiles at $r_{-2}$), host-to-satellite density ratio ($\delta_{\rho}=\rho_{s}(a_{s})/\rho_{h}(r_{c})$), apo-to-pericentre ratio ($r_{a/p}$), apocentre ($r_{a}$), accretion redshift ($z_{i}$) and time ($t_{i}$), cusp regrowth (denoted by yes or no) and sinking timescales $t_{s}$. When the cusp regrows but that oscillations occur before they settle towards a clear answer for the central slope of the density profile, we mark these instances by a y* and give both the time when the subhalo reaches the central region of the host and the time when the oscillations damped out.}
\end{table*}

\section{Results}
We record the timescales for cusp regeneration in Table~1 and present the final density profiles of some N-body models in Figure~2. We find that 1:3 mergers (simulation 3.4.) regrow DM cusps on a relatively short time-scale, $t_{s}\sim 2 \rm{Gyrs}$. Given that these accretion events tend to happen between $z\sim 4-2$ they may not necessarily constitute a threat for the survival of cores, as they leave an ample window of $8-10$ Gyrs for star formation and supernova (SN) feedback to turn the situation around again. However, given that dwarfs like Sculptor (which shows evidence for cored DM profile) have had 90 \% of their star formation prior to 10-11 Gyr ago \citep{Weisz2014}, such mergers may still be problematic. For a mass ratio 1:6 dynamical friction becomes less efficient, so that the cusp reforms at later times, $t_{s}\sim 5 \rm{Gyrs}$ (simulations 1.2.). For mass ratios below 1:8 the accretion of subhalos on the mean of the $c(M,z)$-relation does not lead to cusp rejuvenation. These systems are not dense enough to reach the core of their host galaxy (simulations 1.3., 4.1. and 5.5.) and deposit their material at the outskirts of the host halo. Re-simulating now the evolution of the same satellites at the high-end of the $c-M$ relation (one and two$-\sigma$ times above the mean), we observe that the merger of subhalos with mass ratios as low as 1:30 do lead to cusp regrowth within the allowed time window. Some of these merger events induce oscillations in the host profile as the substructure sinks towards the centre of the halo (simulations 1.2.3s, 1.3.2s, 1.3.3s, 5.5.2s and 5.5.3s). This behaviour has already been reported by \cite{Dekel2003}. The fluctuations eventually die out as the subhalo gets completely disrupted within its host. In those instances we mark the time at which the subhalo first settle at the centre and the profile reaches a convergent solution. 

We note that the shape of the final profile depends primarily on mass ratio, eccentricity and satellite-to-host density ratio $\delta_{\rho}=\rho_{s}(a_{s})/\rho_{h}(r_{c})$. This suggests the existence of a manifold in this parameter space which allow for the regrowth of DM cusps. Delineating its surface would require an extensive set of simulations, which goes beyond the scope of this paper. However, from the results tabulated in Table 1 we can already conclude that only satellites with a density ratio above 0.35 are sufficiently dense to lead to cusp rejuvenation. While such dense subhalos do not dominate in number, their merging timescales can be considerably long ($5-11$ Gyrs). The accretion of these substructures can thus lead to cusp regrowth at relatively low redshifts $z<0.5$. It remains theoretically unclear whether supernova feedback can turn cusps back into cores at such late times given the energetic constraints \citep{Penarrubia2012}. In this respect, accurate measurements of the star formation histories of dwarf galaxies will be key to limit the effects of supernova feedback on the DM distribution in those systems \citep{Tolstoy2009, Weisz2014}. 

The presence of dense, low-mass mostly dark substructures expected to linger in CDM haloes thus hangs like a ``sword of  Damocles'' over the fate of the central density profiles of dwarf galaxies. These systems add a considerable amount of stochasticity to the predicted structure of the DM haloes of dwarf galaxies.
Deriving the likelihood of this type of mergers in cosmological hydrodynamical simulations is an interesting question indeed, but one that requires a large suit of N-body realisations, as the Damocles subhalos constitute a minority within the substructure population. Our N-body experiments suggest that variations in the central density slope of galactic haloes acted on by baryonic feedback are to be expected within the CDM paradigm. If such variations are not found, this could provide considerable support to alternative DM models such self-interacting (SIDM) dark matter \citep{Vogelsberger2012}. \footnote{Due to thermal motions, WDM models should predict cores but simulations show that their sizes are smaller to those inferred by observations \citep{Lovell2014}.} Finally, it is worth noting that the MW's tidal field may strip a large fraction of the substructure population in dwarf galaxy haloes, especially if the accretion of these galaxies onto the MW occurs early on (e.g. $z\sim 2$). Although this may alleviate the Damocles problem described above, we found it to be irrelevant for the haloes considered in this work.
\begin{figure}
\includegraphics[width=0.5\textwidth,trim=0mm 0mm 0mm 0mm,clip]{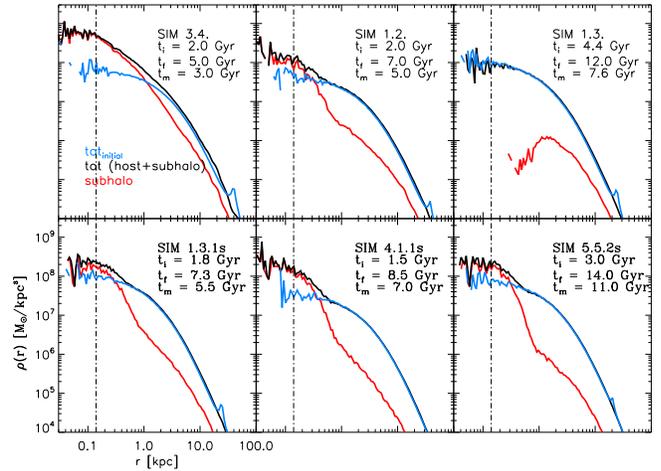}
\caption{Density profiles for some of our realisations: total density profile (black), satellites' particles (red) and total initial density profiles (blue). Simulations runs are denoted as in text and Table 1.  We mark the time of accretion $t_{i}$, the current time $t_{m}$ for which we present the density profiles and the actual corresponding time after accretion $t_{f}=t_{i}+t_{f}$. The vertical dashed line marks the scale 2.8$\epsilon$ above which the force is Newtonian.}
\end{figure}
\section{Discussion and Conclusion}
We have presented a series of collisionless N-body experiments that follow the accretion of massive DM substructure onto MW-type dwarf galaxies after the end of re-ionization ($z<4$). It has been suggested that supernova feedback can under plausible circumstances remove the DM cusps in low-mass haloes, a process that is expected to be most efficient at high redshifts. Here we show that within CDM there may be a cusp regrowth problem. Given the  mass-concentration relation of CDM haloes we have identified cases where cusp regrowth may be expected to happen even at $z\approx 0$. Our experiments suggest the existence of a manifold in the space of the mass-ratio, eccentricity and satellite-to-halo density ratio $\delta_{\rho_{s/h}}$ that allows for cusp regrowth in dwarf galaxies. Delineating this space requires a large suit of cosmological N-body simulations, which is beyond the scope of the current contribution. We note, however, that substructures with a mass ratio above 1:30 and relatively high densities, $\delta_{\rho_{s/h}}>0.35$, always lead to cusp regrowth. Although our models adopt a cuspy profile, this is an ad hoc assumption which remains poorly known from a theoretical and observational point of view. It is also unclear whether these substructures bring in significant amount of stars \citep{Amorisco2014} at infall or remain unaffected by baryons. Baryonic feedback corresponds to another important source of uncertainty in the definition the manifold. Star formation in real dwarves shows a surprisingly rich diversity (see \cite{Weisz2014} for a detailed compilation, or \cite{Kauffmann2014} for statistics at the higher-mass end of dwarves), with some dwarves experiencing most of their star formation at $z\ga 2-1$ (e.g. Sculptor, Cetus or Carina \citep{Weisz2014, Monelli2010, Hurely1998}), while in others (e.g. Leo A \citep{Cole2007}) form the bulk of star formation below $z\la 0.5$. Whether star formation can provide enough feedback energy at low redshifts ($z\la 0.5$) in order to turn DM cusps (such as those found in our experiments) back into cores again without violating the constraints derived from stellar ages is still unclear. Ultimately, the effects studied in this letter need to be assessed in a full cosmological context, although analytical consideration may still prove useful \citep{Penarrubia2012} given that baryonic feedback processes remain in the `sub-grid physics' domain.

According to the reported mass-concentration relation of CDM haloes, the Damocles problem identified in this work mostly arises at the high-density end of the scatter (that is, for subhalos with higher $\rho(r_{s})$ than average). The merging timescales of dense, low-mass subhalos are long enough $5-11~\rm{Gyrs}$ so that depending on the accretion time they can lead to final cusp regrowth by $z=0$. In contrast, we find that the infall of massive substructures (mass ratios above 1:6) and densities on the mean of the relation always leads to cusp regrowth, which may be an issue in the case of the Sculptor dSph given its measured SFH \citep{Weisz2014}. Between 1:6 to 1:8 these need to be above the $1-\sigma$ above the mean, constituting the 10 percentile of the subhalo population. Below 1:8 the chances for regrowth occurs for subhalos above $2-\sigma$ thus a percent of the population.  Our experiments suggest that the central DM slopes of dwarf galaxies acted on by baryonic feedback should show a non-negligible scatter. If such variations are not present in nature and homogeneous density distributions continue to be preferred by observations, this may give strong support to alternative DM particle models such as self-interacting DM, which predict cored density profiles at the low-mass end of the galaxy luminosity function (e.g. \cite{Vogelsberger2012}). On the other hand, a scattered range of halo profiles in dwarf galaxies may provide indirect support for the existence of DM substructures as well as baryonic feedback playing an important role in reshaping the distribution of DM in those galaxies. Measuring the mass profiles of a statistically significant sample of dwarf galaxies appears therefore as a promising avenue to constrain the nature of DM. \section*{Acknowledgments}
CL thanks Benjamin Moster for providing his script for orbit setting.  We thank Matt Walker and Simon White for exciting discussions. CL was supported by a stipend from the MPA and a Junior Fellow award from the Simons Foundation.

\bibliographystyle{mn2e}
\bibliography{master2.bib}{}

\label{lastpage}
\end{document}